\documentclass[a4paper,fleqn]{cas-dc}
\usepackage[numbers]{natbib}
\usepackage{lettrine}
\usepackage{subfigure}
\usepackage{amssymb}
\usepackage{xcolor}
\usepackage{orcidlink}
\usepackage[switch]{lineno}

\def\tsc#1{\csdef{#1}{\textsc{\lowercase{#1}}\xspace}}
\tsc{WGM}
\tsc{QE}

\begin{document}
\let\WriteBookmarks\relax
\def\floatpagepagefraction{1}
\def\textpagefraction{.001}
\let\printorcid\relax 

\shorttitle{}    

\shortauthors{Ziyue Yang et al.}

\title[mode = title]{CSSSTN: A Class-sensitive Subject-to-subject Semantic Style Transfer Network for EEG Classification in RSVP Tasks}

\author[label1]{Ziyue Yang~\orcidlink{0000-0001-6185-5019}}
\author[label1]{Chengrui Chen~\orcidlink{0009-0002-2915-5877}}
\author[label1]{Yong Peng~\orcidlink{0000-0003-1208-972X}}
\author[label2]{Qiong Chen~\orcidlink{0009-0009-5993-8347}}
\author[label1]{Wanzeng Kong\corref{mycorrespondingauthor}~\orcidlink{0000-0002-0113-6968}}
\cortext[mycorrespondingauthor]{Corresponding author}
\ead{kongwanzeng@hdu.edu.cn}
\address[label1]{School of Computer Science and Technology, Hangzhou Dianzi University, Hangzhou 310018, China.}
\address[label2]{Hangzhou Hikvision Digital Technology Company, China, 310051}

\begin{abstract}
The Rapid Serial Visual Presentation (RSVP) paradigm represents a promising application of electroencephalography (EEG) in Brain-Computer Interface (BCI) systems. However, cross-subject variability remains a critical challenge, particularly for BCI-illiterate users who struggle to effectively interact with these systems. To address this issue, we propose the Class-Sensitive Subject-to-Subject Semantic Style Transfer Network (CSSSTN), which incorporates a class-sensitive approach to align feature distributions between golden subjects (BCI experts) and target (BCI-illiterate) users on a class-by-class basis. Building on the SSSTN framework, CSSSTN incorporates three key components: (1) subject-specific classifier training, (2) a unique style loss to transfer class-discriminative features while preserving semantic information through a modified content loss, and (3) an ensemble approach to integrate predictions from both source and target domains. We evaluated CSSSTN using both a publicly available dataset and a self-collected dataset. Experimental results demonstrate that CSSSTN outperforms state-of-the-art methods, achieving mean balanced accuracy improvements of 6.4\% on the Tsinghua dataset and 3.5\% on the HDU dataset, with notable benefits for BCI-illiterate users. Ablation studies confirm the effectiveness of each component, particularly the class-sensitive transfer and the use of lower-layer features, which enhance transfer performance and mitigate negative transfer. Additionally, CSSSTN achieves competitive results with minimal target data, reducing calibration time and effort. These findings highlight the practical potential of CSSSTN for real-world BCI applications, offering a robust and scalable solution to improve the performance of BCI-illiterate users while minimizing reliance on extensive training data. Our code is available at https://github.com/ziyuey/CSSSTN.
\end{abstract}




\maketitle

\section{Introduction}
A Brain-Computer Interface (BCI) enables direct communication between the brain and external devices, bypassing traditional body functions \cite{he2020brain, kawala2021summary}. BCIs have diverse applications, including helping patients with brain injuries, supporting complex tasks such as driving and surgery, and enabling regulation of brain activity for therapeutic purposes (e.g., treating depression) \cite{hramov2021physical}. Scalp electroencephalography (EEG) is widely used in BCI research due to its non-invasive nature, high temporal resolution, portability, and low cost \cite{islam2023recent, janapati2023advances}.

One of the most promising applications of EEG in BCI systems is the use of Rapid Serial Visual Presentation (RSVP) paradigms, which involve the presentation of visual stimuli at a frequency of 5–20 Hz \cite{bigdely2008brain,lees2018review}. During these visual presentations, information related to objects of interest can trigger the generation of an event-related potential \cite{lees2019speed}. The P300 component, a positive deflection in the EEG occurring 250 to 500 ms after a target stimulus, is the ERP most commonly used in RSVP-BCI applications. It is closely related to attention and memory processes, helping to identify the subject's focus and detect the most relevant stimuli \cite{luo2010three,lees2018review}.

The combination of RSVP paradigms with advanced EEG decoding techniques has led to the development of highly efficient automated image processing systems \cite{matran2016brain,wu2024semantic,li2021enhancing}. RSVP-based BCI systems are capable of faster detection and recognition of relevant objects and information compared to traditional manual analysis, significantly improving the efficiency of professionals \cite{lees2018review}. These systems are particularly useful in fields such as counterintelligence, law enforcement, and healthcare, where professionals are tasked with reviewing large volumes of images or data. In addition, RSVP-BCIs are being explored for security and authentication applications, where the P300 signal serves as a biometric marker for identity verification, offering a secure and non-invasive method of authentication \cite{zeng2018eeg}. 

Despite the great potential of RSVP-based BCIs, their transition from laboratory settings to real-world applications faces several significant challenges, which are cross-subject, cross-time, and cross-scene \cite{wan2021review}. These issues are not unique to RSVP-based systems but are prevalent across various EEG-based BCI applications (e.g., motor imagery). In real world settings, EEG data from different users and sessions can vary substantially due to intra- and intersubject variability \cite{saha2020intra} that involves individual differences in brain activity patterns, physiological noise level \cite{kappel2017physiological}, signal quality \cite{raduntz2018signal}, and even emotional states \cite{zhao2021plug}. This variability presents a considerable challenge in constructing robust classifiers that can be generalized between different subjects and sessions. A particularly serious consequence of this variability is the phenomenon known as BCI illiteracy \cite{lee2019eeg}, which refers to the difficulty or inability of some users to effectively control or interact with BCI systems, despite the capabilities of the technology. In critical applications, these problems can severely limit the effectiveness and applicability of technology. 

To address these challenges, several approaches based on machine learning and deep learning have been proposed. However, these methods still face some disadvantages. First, domain shifts caused by intra and intersubject variability make it difficult to identify common domain-invariant feature representations in multiple subjects \cite{saha2020intra}. As a result, traditional domain adaptation techniques may not extract meaningful features from the source domain due to the large discrepancies between subjects. Second, these approaches may suffer from negative transfer, which occurs when knowledge transferred from the source domain negatively impacts the classifier’s performance on the target domain instead of improving it \cite{jimenez2021study}. This is particularly problematic when extracting common representations from subjects with vastly different brain activity patterns, leading to poor generalization. To mitigate these limitations, recent research on domain adaptation has explored style transfer techniques, which have shown promise, particularly in motor imagery applications \cite{kim2023bridging, sun2022golden}. In these methods, a subject-style transfer neural network has been proposed to transform the data distribution of BCI-illiterate subjects into that of BCI-expert subjects, often referred to as "golden subjects." Despite their potential, style transfer techniques have not yet been explored in RSVP-based BCIs, where the class imbalance problem is particularly prominent. Moreover, existing style transfer methods tend to overlook class-specific features during the transfer process, which may limit their overall effectiveness.

Therefore, this study proposes a class-sensitive subject-to-subject semantic style transfer network (CSSSTN) to address the cross-subject variability and BCI illiteracy problem in RSVP-based BCIs.The performance of the proposed method is thoroughly evaluated on both a publicly available dataset and a self-collected dataset. The key contributions of our proposed method are as follows:
\begin{itemize} 
\item Introduction of Subject-to-Subject Semantic Style Transfer: We adapt subject-to-subject style transfer to RSVP-based BCIs to improve cross-subject knowledge transfer.
\end{itemize}

\begin{itemize} 
\item Class-Sensitive Transfer: We propose a class-sensitive approach where the style loss aligns feature distributions for each class separately, preserving class-specific information and improving transfer effectiveness.
\end{itemize}

\begin{itemize} 
\item Impact of "Golden Subject" and Sample Selection: We discuss how the choice of the "golden subject" and the selection of training samples affect the overall quality of the style transfer process.
\end{itemize}

\section{Related work}
To improve BCI illiteracy and improve the performance and reliability of EEG-based BCIs, various machine learning methods have been developed. Some approaches focus on the user-BCI interaction, such as Vidaurre et al.'s co-adaptive learning using linear discriminant analysis, which reduces cross-subject performance variations through closed-loop feedback \cite{vidaurre2010towards}. Other research has concentrated on feature space alignment. For example, Wu et al. proposed a method to align EEG trials from different subjects in Euclidean space \cite{he2019transfer}, while Tao et al. introduced a multi-kernel learning approach that aims to minimize feature distribution discrepancies and enhance class separability \cite{tao2022distribution}. As research progressed, deep learning and domain adaptation methods gained attention for their ability to extract common features across subjects. Li et al. proposed a multisource transfer learning method for EEG emotion recognition \cite{li2019multisource}, and Zhao et al. introduced a deep representation-based domain adaptation (DRDA) method to leverage domain-invariant features \cite{zhao2020deep}. Jeon et al. extended this by using mutual information to refine feature selection \cite{jeon2021mutual}. Hang et al. introduced a deep domain adaptation network (DDAN) that minimizes feature distribution discrepancies using the maximum mean discrepancy and improves classification accuracy across subjects \cite{hang2019cross}.

In the context of RSVP-based BCIs, several studies have focused on improving cross-subject classification. Liu et al. proposed the Correlation Analysis Rank (CAR) algorithm, which improves performance by sorting the correlation between subjects, outperforming traditional random selection methods \cite{liu2020improving}. Wang et al. introduced a multi-source domain adaptation-based tempo-spatial convolution network (MDA-TSC) to align feature distributions across subjects \cite{wang2024multi}, while Zhang et al. developed a multilevel information fusion model to enhance EEG stability in dual-subject RSVP tasks \cite{zhang2021two}.

Style transfer, originally applied in computer vision, has recently been adapted to classification of motor imagery based on EEG with promising results \cite{sun2022golden, kim2023bridging}. Sun et al. proposed a subject transfer neural network (STNN) that directly transforms the data distribution of BCI-illiterate subjects into golden subjects \cite{sun2022golden}. Building on this, Kim et al. further developed a subject-to-subject semantic style transfer network (SSSTN) that preserves content information from the target domain while transferring the style from the source domain \cite{kim2023bridging}. However, SSSTN methods have some limitations. A major issue is their overlook of class-specific information during the transfer process, which can be particularly problematic in cases of class imbalance or substantial class discrepancies. Another limitation is the reliance on the entire data set from the target subject without discussing the impact of the data set size on transfer performance. This could potentially increase the time and effort required for data collection. Additionally, SSSTN employs features from all convolutional layers during the transfer process, which may limit flexibility and increase computational time. To address these challenges, this study introduces style transfer for cross-subject RSVP detection and proposes extensions to SSSTN, which significantly improve the performance of class-specific transfer.

\section{Experimental setup}
\subsection{Tsinghua RSVP Dataset}
The dataset was provided by the Tsinghua Brain-Computer Interface Research Group and contains EEG data from 64 healthy subjects \cite{zhang2020benchmark}.They recorded the EEG data using a 64-channel EASYCAP with electrodes arranged in accordance with the standard 10–10 system\cite{nuwer1998ifcn}, and a Brainvision actiCHamp amplifier. Each subject was instructed to sit in front of a screen and participate in a RSVP task, where the images were presented at a frequency of 10 Hz (10 images per second). The stimulus images were sourced from the Massachusetts Institute of Technology Computer Science and Artificial Intelligence Library. These images consisted of two categories: target images containing humans and non-target images without humans. EEG signals were recorded at a sampling rate of 1000 Hz. The goal was to identify whether a given image was a target or nontarget based on the recorded EEG signals. For this study, EEG data from the first 10 subjects were selected to evaluate the proposed model. Additionally, non-target data were downsampled to maintain a 10:1 ratio between non-target and target samples.
\subsection{HDU RSVP Dataset}
\begin{figure}[t]
  \centering
   \includegraphics[width=0.95\linewidth]{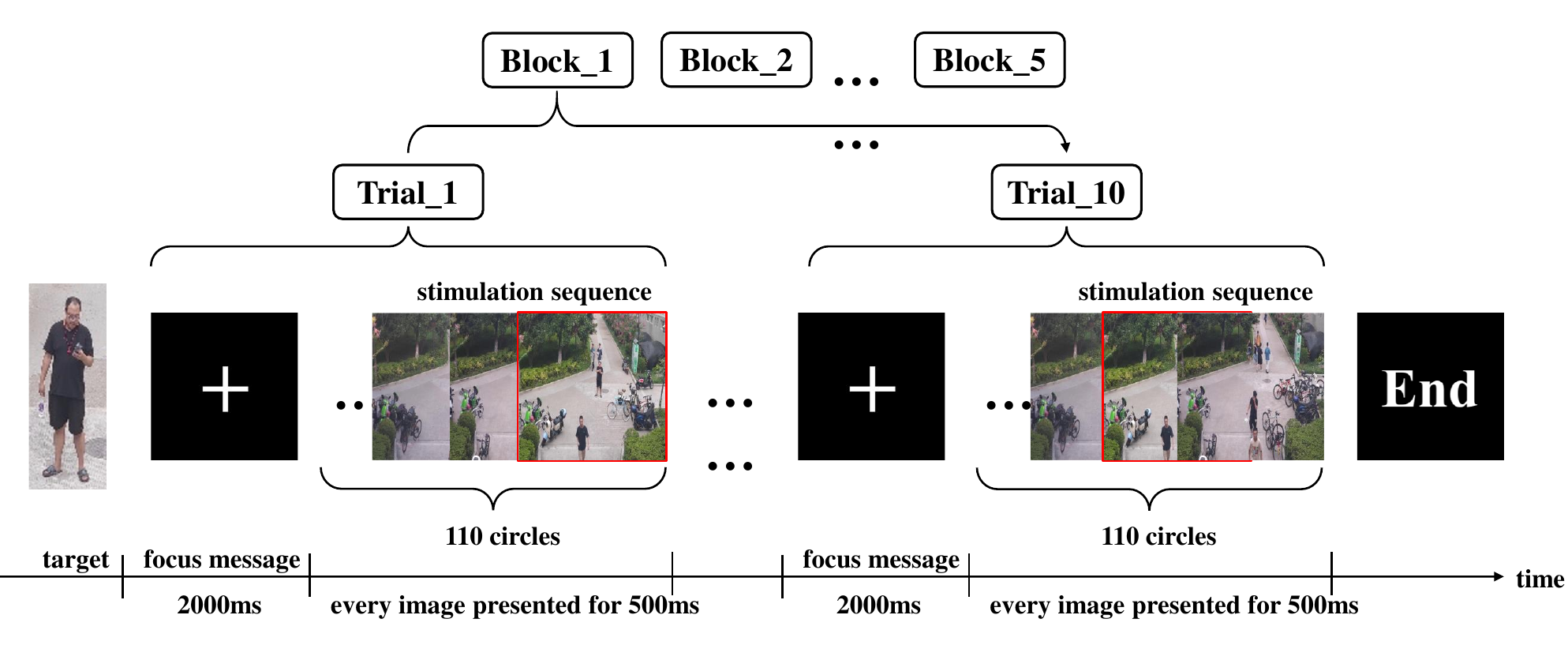}
   \caption{Experimental scheme in which subjects identify target images within a RSVP sequence, generating the corresponding EEG signals.}
   \label{fig:Experimental scheme}
\end{figure}

We collected a self-constructed dataset involving 10 right-handed participants (all men). All participants had normal or corrected normal vision and none reported any psychiatric disorders or relevant family medical history. The study was reviewed and approved by the
Ethics Committee of Second Affiliated Hospital of Zhejiang University, College of Medicine and the protocol number is IRB-2024-1535. Signed informed consent was obtained from each participant.
The electroencephalogram (EEG) acquisition subsystem consists of a 64-channel QuikCap EEG cap from Neuroscan (Australia) and SynAmps2 amplifiers, with Scan 4.5 software being utilized for data recording and processing.The 64-channel QuikCap EEG electrode array was arranged in accordance with the international 10-20 system for standardized electrode placement\cite{jasper1958ten}.The stimulus material for this RSVP-based target detection experiment was derived from a self-collected video data set recorded using two 4K cameras. The images were extracted at a sampling rate of one frame per second and a pedestrian was selected as the target task. The participants sat in a quiet and comfortable environment and were instructed to identify images containing the target stimulus from a sequence of rapidly presented images. The experimental paradigm, process, and parameter settings are shown in Fig.~\ref{fig:Experimental scheme}. The experiment consisted of 8 blocks, each block containing 5 trials. Each trial included 110 stimulus images, with a target-to-nontarget image ratio of 1:10. Each participant completed the experiment five times. Before starting each experiment, participants were shown an image containing the target as a cue. To help participants focus, a fixation cross was displayed in the center of the screen for 2000 ms at the start of each trial. The stimulus images were then presented at a rate of 2 Hz. An image of the end screen indicated the conclusion of each trial. Participants were allowed to take breaks of any duration between blocks and resumed the experiment at their discretion. To ensure variety, at least one non-target image was presented between any two target images. The experiment was carried out using E-Prime 3.0 for the presentation of the stimulus. All stimulus images were displayed on a monitor with a resolution of 1680 × 1050 and a refresh rate of 60 Hz.

\subsection{Data Preprocessing}
The raw EEG data were preprocessed using the MNE software package, including filtering and baseline correction. The sampling rate was reduced to 250~Hz, and a 2--30~Hz FIR band-pass filter was applied to remove power line interference and high-frequency noise. EEG data were segmented into 1-second epochs starting from stimulus onset, resulting in 2200 samples per subject, each with dimensions $ C \times 250 $ ($ C $ is the number of channels).

To extract time-frequency features, the continuous wavelet transform (CWT) was applied. Compared to the Fourier Transform, CWT better captures time-varying, non-stationary signals, which is crucial for weak ERP components like P300. The Morlet wavelet, defined as:

\begin{eqnarray}
    \psi(t) & = & \exp\left(-\frac{\beta^2 t^2}{2}\right)\cos(\pi t),
\end{eqnarray}

was used for its excellent time-frequency resolution. 

The resulting time-frequency maps were cropped to \( 28 \times 100 \) and upsampled to \( 64 \times 64 \) using bilinear interpolation. This process was applied in all channels, producing a final feature matrix of size \( C \times 64 \times 64 \).
\subsection{Competing Methods}
We evaluated the proposed CSSSTN by comparing its performance with the following competing methods:

\begin{itemize}
    \item \textbf{EEGNet}: A compact convolutional neural network that uses depthwise and separable convolutions for effective classification of EEG signals across different BCI paradigms \cite{lawhern2018eegnet}.
    \item \textbf{XDAWN-Riemann}: Combines the XDAWN spatial filter with the Minimum Distance to Riemannian Mean classifier. The Riemannian geometry framework enhances EEG decoding performance compared to Euclidean metrics \cite{delgado2020riemann}.
    \item \textbf{DeepConvNet}: A deep convolutional neural network designed for raw EEG data, utilizing hierarchical feature extraction via multiple convolutional and pooling layers to improve classification accuracy \cite{schirrmeister2017deep}.
\item \textbf{CNN}: A CNN classifier based on CWT data, designed to extract time-frequency features. This model serves as the base classifier for both CSSSTN and STNN (see Fig.~\ref{fig:classifier}).
    \item \textbf{SE-CNN}: Extends CNN by incorporating a squeeze-and-excitation module to capture inter-channel relationships. This model is used by SSTN for EEG classification \cite{zhang2021motor}.
\item \textbf{STNN}: Utilizes a golden subject template to enhance the performance of BCI-illiterate users. It employs BCE loss and perceptual loss during training to improve transfer learning \cite{sun2022golden}.
\item \textbf{SSSTN}: Leverages style loss, content loss, and perceptual loss to enhance classification performance, demonstrating excellent results in the motor imagery domain \cite{kim2023bridging}.
\end{itemize}

The results were assessed using balanced accuracy, which is defined as
\begin{eqnarray}
    \text{Balanced Accuracy} & = & \frac{\frac{\text{TP}}{\text{TP} + \text{FN}} + \frac{\text{TN}}{\text{TN} + \text{FP}}}{2},
\end{eqnarray}
where \( \text{TP} \), \( \text{FN} \), \( \text{TN} \), and \( \text{FP} \) represent true positives, false negatives, true negatives, and false positives, respectively. Balanced accuracy provides a robust evaluation metric for unbalanced datasets by equally weighting the performance in both positive and negative classes.

\section{Methods}
\begin{figure*}[ht]
    \centering
    \includegraphics[width=\textwidth]{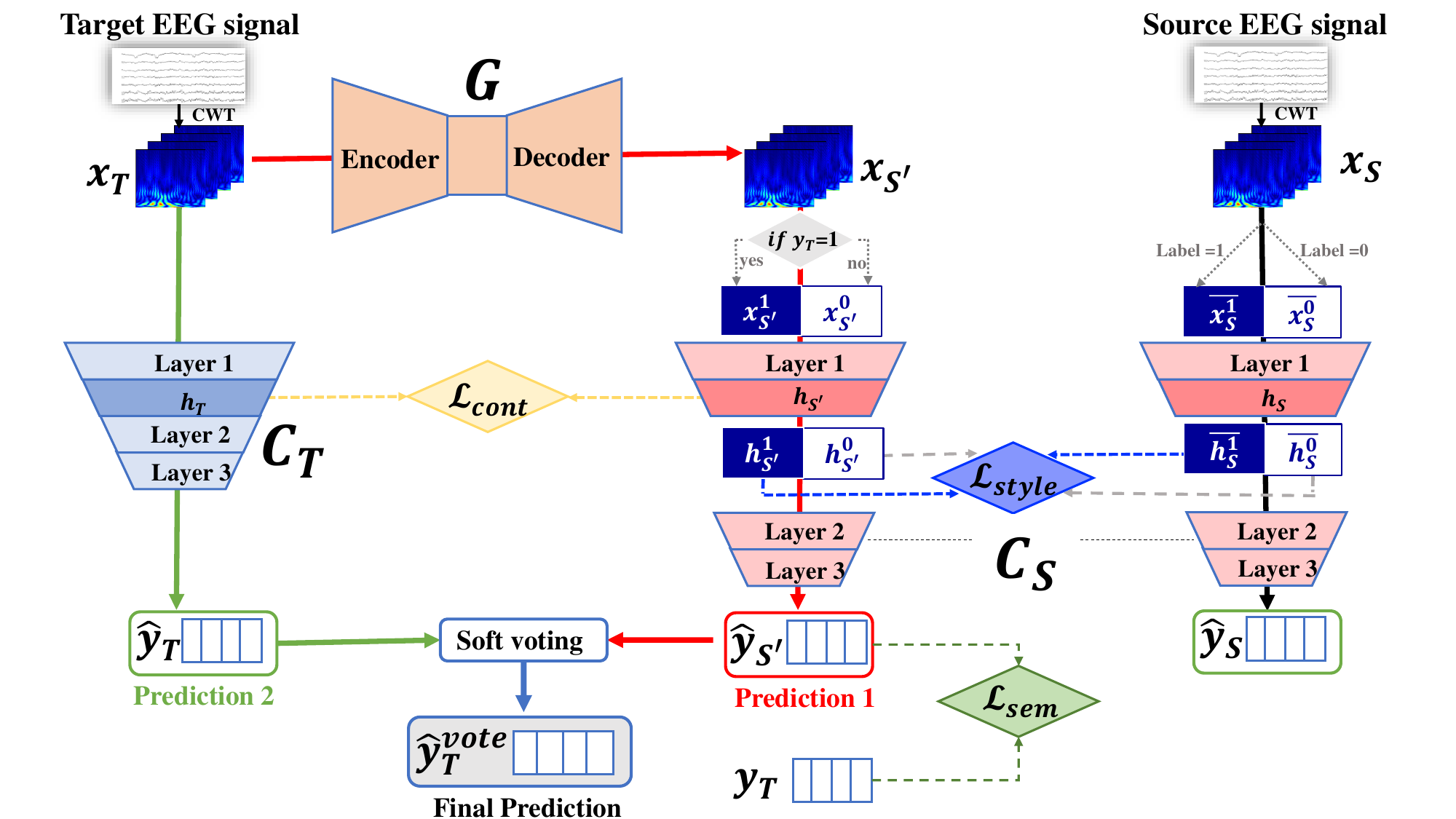} 
\caption{Overview of the proposed CSSSTN framework. The framework consists of three phases: (1) pretraining, (2) style transfer, and (3) prediction and ensemble.  
Input data \( x_T \) (target subject) and \( x_S \) (source subject) are used, while \( x_S' \) is the source data transformed by generator \( G \). In the pretraining phase, classifiers \( C_T \) and \( C_S \) are trained on \( x_T \) and \( x_S \), respectively.  During the style transfer phase, content loss \( L_{\text{cont}} \) is computed between first-layer features \( h_{T} \) and \( h_{S'} \), extracted from \( x_T \) and \( x_S' \) by \( C_T \) and \( C_S \). Style loss \( L_{\text{style}} \) is computed to align the target-transformed features \( h_{S'} \) with source class templates \( \bar{h}_{S}^0 \) and \( \bar{h}_{S}^1 \), which are the averaged features of \( x_S \) for each class (non-target and target). Depending on \( y_T \) (the target label), style loss is calculated as the KL divergence between \( h_{S'}^0 \) and \( \bar{h}_{S}^0 \), or between \( h_{S'}^1 \) and \( \bar{h}_{S}^1 \).  Semantic loss \( L_{\text{sem}} \) ensures that the predicted label \( \hat{y}_{S'} \) (from \( C_S \)) matches the ground-truth \( y_T \). The final prediction is obtained using a soft voting ensemble of \( \hat{y}_{S'} \) and \( \hat{y}_T \) (from \( C_T \)).}

    \label{fig:Overview}
\end{figure*}

We propose CSSSTN to address cross-subject stability and BCI illiteracy in RSVP-EEG paradigms. EEG signals are converted into time-frequency features using CWT. To handle the low signal-to-noise ratio and the class imbalance issue, we constructed an optimal template for each class, helping the model focus on class-specific features during transfer. The experimental design is shown in Fig.~\ref{fig:Overview}.
\subsection{Pretraining}
\begin{figure}[t]
  \centering
   \includegraphics[width=0.95\linewidth]{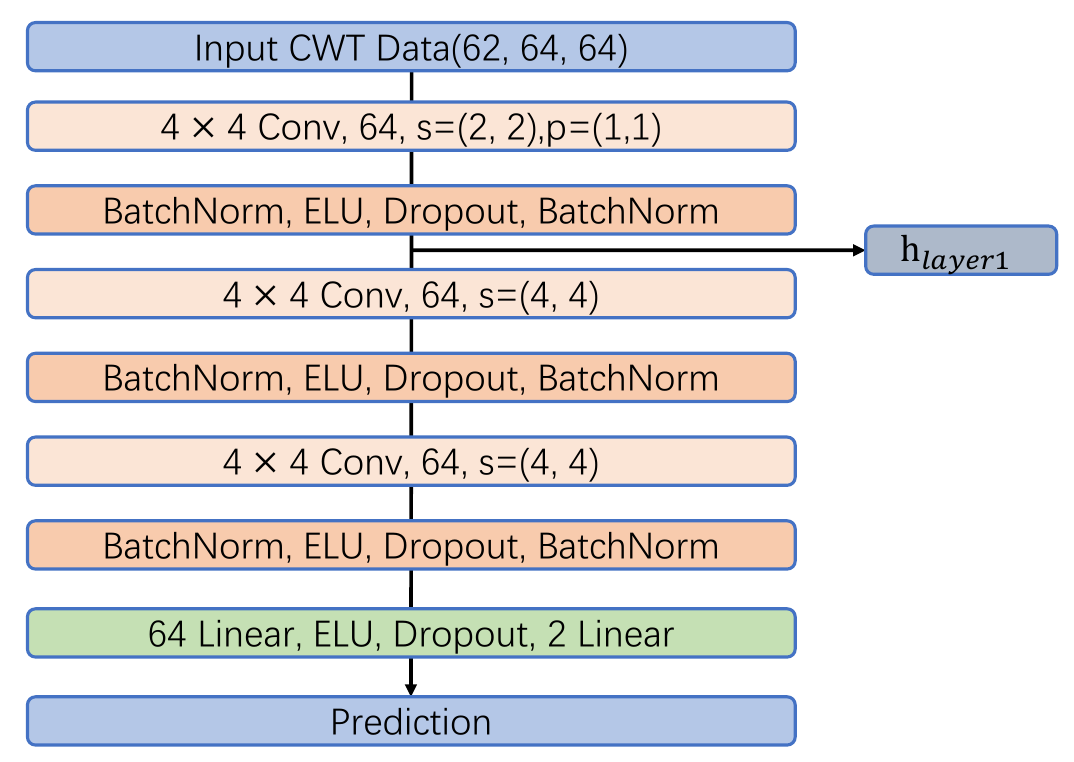}
   \caption{Architecture of the classifier \(\mathbf{C}\) in the proposed framework. Each convolutional layer specifies the kernel size and the number of output channels. Variables \(s\) and \(p\) denote the stride and padding, respectively. The activation function used is ELU. The output features of the first convolutional blocks \(h_{\text{layer1}}\) is utilized for loss computation. The linear layers (e.g., 64 and 2) indicate the output dimensions, with the final prediction size being 2.}
   \label{fig:classifier}
\end{figure}
At this stage, CNN binary classifiers are trained for each subject using the precomputed time-frequency feature matrix.  The classifiers consist of three 2D convolutional blocks for feature extraction and one fully connected block for classification, as shown in Fig.~\ref{fig:classifier}. Each convolutional block extracts time-frequency features within its receptive field, while the ELU activation function ensures robustness by allowing non-zero outputs for negative inputs. Dropout is introduced to accelerate training and reduce overfitting.  

The output of the first convolutional blocks is saved as lower-level features for later use in the feature transfer model. Given the class imbalance in RSVP tasks (target to non-target ratio of 1:10), the loss function incorporates class weights based on the inverse frequency of each class:
\begin{eqnarray}
    L_{\text{cla}} &=& -\sum_{k=1}^K \omega_k y^{(k)} \log(\hat{y}^k),
\end{eqnarray}

where \( K \) is the number of classes, \( \omega_1 = 1 \) (non-target) and \( \omega_2 = 10 \) (target). After training, classifiers achieving the best performance are selected as source subjects, while others are used as target subjects by default. 

\subsection{Style Transfer}
\begin{figure}[t]
  \centering
   \includegraphics[width=0.95\linewidth]{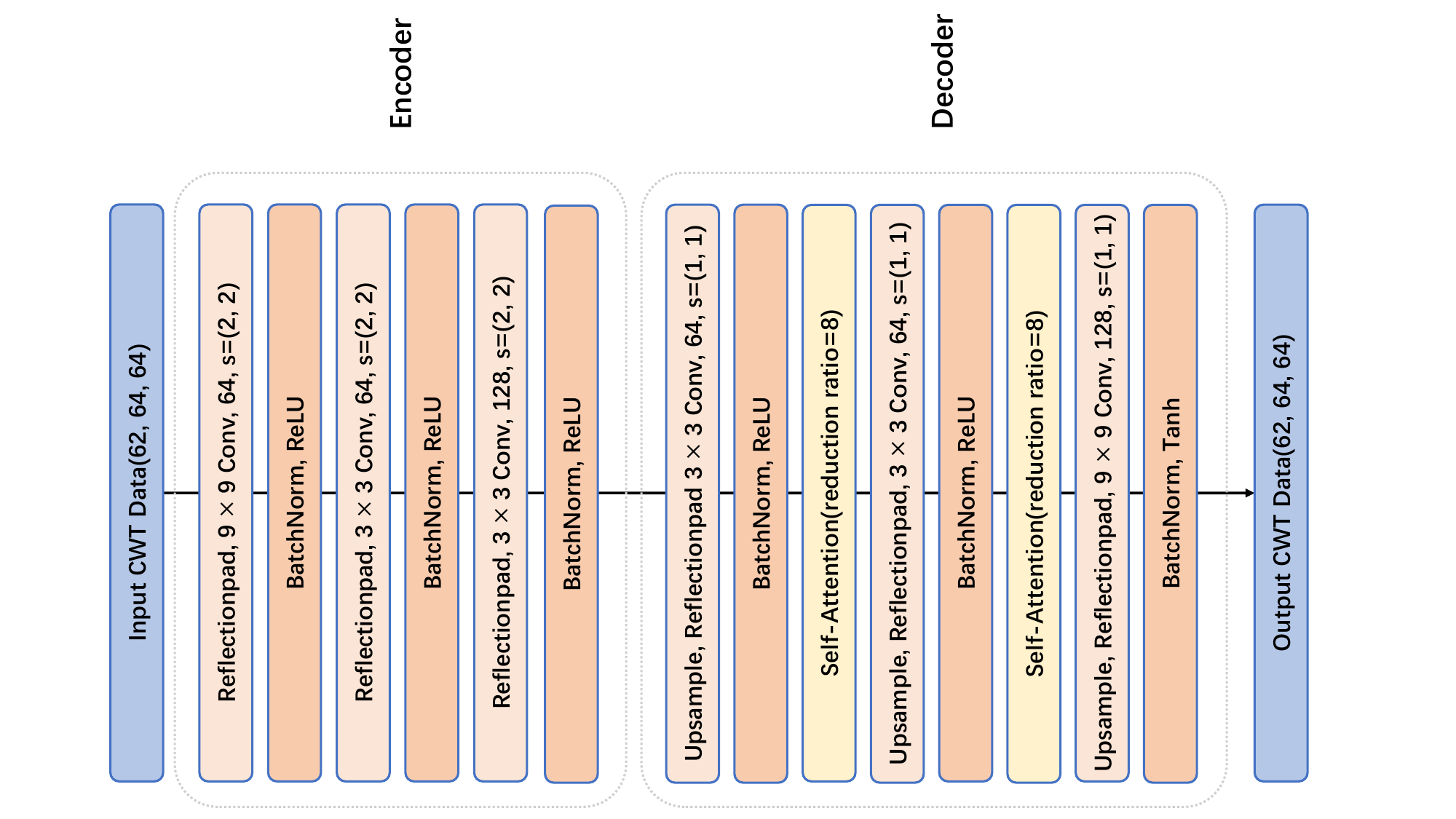}
   \caption{Architecture of the generator \(\mathbf{G}\) in the proposed framework, using an encoder-decoder structure with self-attention modules to enhance feature representation. The encoder compresses the input via convolutional layers, while the decoder reconstructs the output with upsampling and self-attention mechanisms, ensuring high-quality reconstruction.}
   \label{fig:generator}
\end{figure}

To address the variability between subjects common in EEG-BCIs applications, we used a framework similar to SSSTN \cite{kim2023bridging} with modifications. The generator \( G \) adopts an auto-encoder structure comprising three convolutional blocks in both the encoder and the decoder (see Fig.~\ref{fig:generator}). Each block includes normalization, dropout, and activation layers. A self-attention module is added after each convolutional block to emphasize key local features by assigning appropriate weights.  

During training, pre-trained classifiers \( C_T \) and \( C_S \) (for target and source subjects, respectively) are used to extract classification features. To handle class-sepcific information, class templates are constructed as:
\begin{eqnarray}
\bar{X_S}^k & = & \frac{1}{|X_{S}^k|} \sum_{x \in X_{S}^k} x,
\end{eqnarray}
where \( X_{S}^k \) is the set of samples belonging to class \( k \), \( |X_{S}^k| \) is the sample count and \( \bar{X_{S}^k} \) is the class template. 

For training \( G \), target samples \( x_T \) are input to the generator to produce \( x_S' \). Both \( x_S' \) and the template of the source class \( \bar{X}_S^k \) are passed through \( C_S \) to obtain the feature representations \( h_{S'} \) and \( \bar{h_{S}^k} \). To ensure that \( x_T \) is effectively mapped to the feature space of the source, the style loss is defined as: \begin{eqnarray}
 L_{\text{style}} & = & \text{KL}(h_{S'} \,||\, \bar{h_{S}^k}), 
\end{eqnarray}
where \(k\) is the class of \( x_T \) and the KL divergence measures the distance between the two distributions.  

To preserve semantic content during transfer, content loss is introduced:
\begin{eqnarray}
L_{\text{cont}} & = & \left( h_{S'}- h_T \right)^2,
\end{eqnarray}
where \( h_{S'} \) and \( h_T \) are characteristics extracted from the third convolutional layer of \( C_S \) and \( C_T \), respectively.  

Finally, a classification loss is added for \( x_S' \) to match its label:
\begin{eqnarray}
    L_{\text{sem}} &=& -\sum_{k=1}^K y^{(k)} \log(\hat{y}^k).
\end{eqnarray}

The total loss function for training \( G \) is:
\begin{eqnarray}
L_{\text{total}} & = & L_{\text{style}} + L_{\text{cont}} + L_{\text{sem}}.
\end{eqnarray}

To address class imbalance during the style transfer phase, target samples are oversampled before training to achieve a balanced target-to-non-target ratio of 1:1. This oversampling approach is applied to our method as well as the STNN and SSSTN methods to ensure fairness in comparison.

\section{Results}
\subsection{Comparison of CSSSTN with Baseline Methods}
We first compared CSSSTN with baseline methods in the TsingHua dataset. The results are presented in Table~\ref{table1}. The average accuracies for EEGNet, XDAWN-riemann, DeepConvNet, CNN, SE-CNN, STNN, SSSTN and CSSSTN were 81.0\%, 82.4\%, 78.8\%, 84.3\%, 84.8\%, 58.3\%, 80.2\%, and 91.2\%, respectively. CSSSTN achieved an average accuracy improvement of 6.4\% over the best state-of-the-art methods and 11.0\% over SSSTN, demonstrating its superior classification capability. In particular, CSSSTN outperformed all baseline methods in all the 10 subjects.

Similar results were observed in the HDU dataset, as shown in Table~\ref{table2}. CSSSTN achieved average accuracy improvements of 22.2\%, 15.1\%, 22.3\%, 11.6\%, 13.1\%, 20.8\%, 3.5\% over EEGNet, XDAWN-riemann, DeepConvNet, CNN, SE-CNN, STNN and SSSTN, respectively, further demonstrating its effectiveness. CSSSTN performed better than SSSTN for all 7 subjects and outperformed other competing methods (except SSSTN) for 9 subjects.

It should be noted that our CSSSTN consistently demonstrated positive transfer effects across both datasets, highlighting the robustness of the proposed method. In contrast, STNN exhibited negative transfer effects in both datasets, with performance dropping by 26\% and 9.2\% compared to CNN. SSSTN exhibited a slight performance drop of 4.6\% on the Tsinghua dataset but performed well on the HDU dataset, achieving a 9.6\% improvement over SE-CNN. However, even on the HDU dataset, SSSTN experienced negative transfer for Subject N10, with performance dropping by 2.7\% compared to SE-CNN. In contrast, CSSSTN demonstrated no negative transfer across both datasets and all subjects, highlighting its robustness.

\begin{table*}[!h]\tiny
    \renewcommand{\arraystretch}{1.2}
    \caption{Mean balance accuracy and standard deviation (std) results for EEGNet, XDAWN-riemann, DeepConvNet, CNN, SE-CNN, STNN, SSSTN, and CSSSTN on TsingHua dataset. * Indicates the golden subject without transfer.}
    \label{table1}
    \centering
    \resizebox{\textwidth}{!}{
        \begin{tabular}{c c c c c c c c c}
            \hline
            \multicolumn{1}{c}{\multirow{2}{*}{Subject}} & \multicolumn{8}{l}{Balance Accuracy \% (mean $\pm$ std)} \\ 
            \cline{2-9} 
            \multicolumn{1}{c}{}   & EEGNet     & XDAWN-riemann    & DeepConvNet    & CNN    & SE-CNN  & STNN  & SSSTN  & CSSSTN(Ours)   \\ 
            \hline
            No.1 & 83.6±1.4 & 84.9±2.2 & 83.9±1.4 & 85.7±3.2 & 87.0±2.0 & 52.1±0.8 & 76.0±1.8 & \textbf{91.7±1.6} \\ 
             No.2 & 74.3±2.5 & 75.0±2.5 & 70.5±2.6 & 76.9±2.0 & 80.9±3.0 & 52.8±1.4 & 82.4±3.8 & \textbf{86.2±1.1} \\ 
             No.3 & 84.6±2.9 & 83.6±3.5 & 83.4±3.2 & 91.2±1.3 & 90.7±2.3 & 61.0±8.2 & 80.5±2.7 & \textbf{94.8±1.8} \\ 
            No.4 & 80.3±5.3 & 78.9±3.9 & 75.1±3.0 & 77.9±2.1 & 80.7±3.9 & 53.3±1.5 & 81.0±4.6 & \textbf{87.1±3.2} \\ 
            No.5 & 82.8±3.7 & 81.2±4.0 & 81.5±4.6 & 86.2±3.2 & 85.3±2.8 & 51.8±0.8 & 87.1±6.9 & \textbf{92.7±3.4} \\ 
            No.6 & 86.6±2.6 & 86.3±3.1 & 85.7±2.0 & 88.5±1.5 & 90.0±2.6 & 61.6±4.4 & 89.7±2.3 & \textbf{91.5±1.9} \\ 
            No.7 & 85.7±2.2 & 86.9±1.5 & 84.8±2.8 & 89.6±1.8 & 87.9±1.4 & 53.5±1.7 & 76.5±2.2 & \textbf{91.3±3.2} \\ 
            No.8 & 79.2±6.5 & 81.4±3.7 & 77.6±4.9 & 83.7±3.6 & 82.7±4.0 & 53.9±1.5 & 72.5±3.2 & \textbf{93.5±4.3} \\ 
            No.9 & 64.7±4.2 & 66.7±4.0 & 59.5±3.5 & 73.1±3.3 & 71.3±3.5 & 51.6±0.7 & 65.6±3.3 & \textbf{91.4±4.7} \\ 
            No.10 & 89.1±1.7 & 89.3±1.5 & 85.6±1.7 & \textbf{91.5±1.3} & {91.2±1.8} & \textbf{91.5±1.3*} & {91.2±1.8*} & \textbf{91.5±1.3*} \\ 
            \cline{1-9} 
            Average & 81.0±3.3 & 82.4±3.0 & 78.8±2.7 & 84.3±5.3 & 84.8±2.7 & 58.3±2.3 & 80.2±3.2 & \textbf{91.2±2.7} \\ 
            \hline
        \end{tabular}
    }\\
\end{table*}

\begin{table*}[!h]\tiny 
    \renewcommand{\arraystretch}{1.2} 
    \caption{Mean balance accuracy and standard deviation (std) results for EEGNet, XDAWN-riemann, DeepConvNet, CNN, SE-CNN, STNN, SSSTN, and CSSSTN on HDU dataset. * Indicates the golden subject without transfer.}
    \label{table2}
    \centering
    \resizebox{\textwidth}{!}{ 
        \begin{tabular}{c c c c c c c c c}
            \hline
            \multicolumn{1}{c}{\multirow{2}{*}{Subject}} & \multicolumn{8}{l}{Balance Accuracy \% (mean ± std)} \\ 
            \cline{2-9} 
            \multicolumn{1}{c}{}   & EEGNet     & XDAWN-riemann    & DeepConvNet    & CNN    & SE-CNN  & STNN  & SSSTN  & CSSSTN(Ours) \\ 
            \hline
             No.1 & 61.5±2.2 & 75.3±4.6 & 60.4±2.9 & 75.7±4.5 & 73.6±4.1 & 69.8±2.2 & 83.7±5.1 & \textbf{92.1±2.7} \\ 
             No.2 & 63.1±4.5 & 71.8±2.5 & 62.0±4.2 & 76.9±1.3 & 70.4±2.2 & 70.0±2.2 & 89.6±8.9 & \textbf{95.0±3.5} \\ 
             No.3 & 76.6±3.6 & \textbf{84.0±1.3} & 74.6±4.1 & 80.8±4.9 & 77.6±4.6 & {80.8±4.9*} & {77.6±4.6*} & {80.8±4.9*} \\ 
             No.4 & 69.9±3.8 & 71.4±3.3 & 70.2±3.0 & 73.3±4.1 & 75.2±4.5 & 60.9±2.2 & 80.5±9.0 & \textbf{86.1±6.7} \\ 
             No.5 & 68.4±4.8 & 77.4±1.1 & 70.4±3.6 & 84.4±1.9 & 81.0±3.7 & 74.4±2.1 & \textbf{96.8±5.4} & 86.8±4.1 \\ 
             No.6 & 72.4±8.2 & 80.3±5.2 & 72.7±9.3 & 82.2±1.1 & 83.6±1.5 & 59.3±1.3 & \textbf{92.6±7.5} & 64.5±5.9 \\ 
             No.7 & 58.1±2.8 & 73.6±3.2 & 56.2±4.1 & 72.8±3.4 & 73.1±2.1 & 61.1±1.1 & 79.3±7.7 & \textbf{90.1±8.2} \\ 
             No.8 & 73.3±3.1 & 60.3±9.9 & 74.8±2.5 & 79.7±1.2 & 76.9±3.0 & 68.1±1.5 & 84.3±5.1 & \textbf{94.6±4.8} \\ 
             No.9 & 54.2±3.7 & 58.4±4.4 & 52.6±1.7 & 59.7±1.9 & 61.2±2.7 & 61.7±1.9 & 86.0±5.8 & \textbf{93.6±3.5} \\ 
             No.10 & 58.6±3.2 & 65.4±4.6 & 56.2±2.7 & 71.6±4.1 & 70.2±3.0 & 58.5±1.7 & 67.5±4.9 & \textbf{89.2±5.9} \\ 
             \cline{1-9} 
             Average & 65.1±3.9 & 71.8±4.0 & 65.0±3.8 & 75.7±2.8 & 74.2±3.1 & 66.5±2.1 & 83.8±6.5 & \textbf{87.3±5.0} \\ 
            \hline
        \end{tabular}
    }\\
\end{table*}

\subsection{TSNE Visualization}
\begin{figure}[t]
  \centering
   \includegraphics[width=0.95\linewidth]{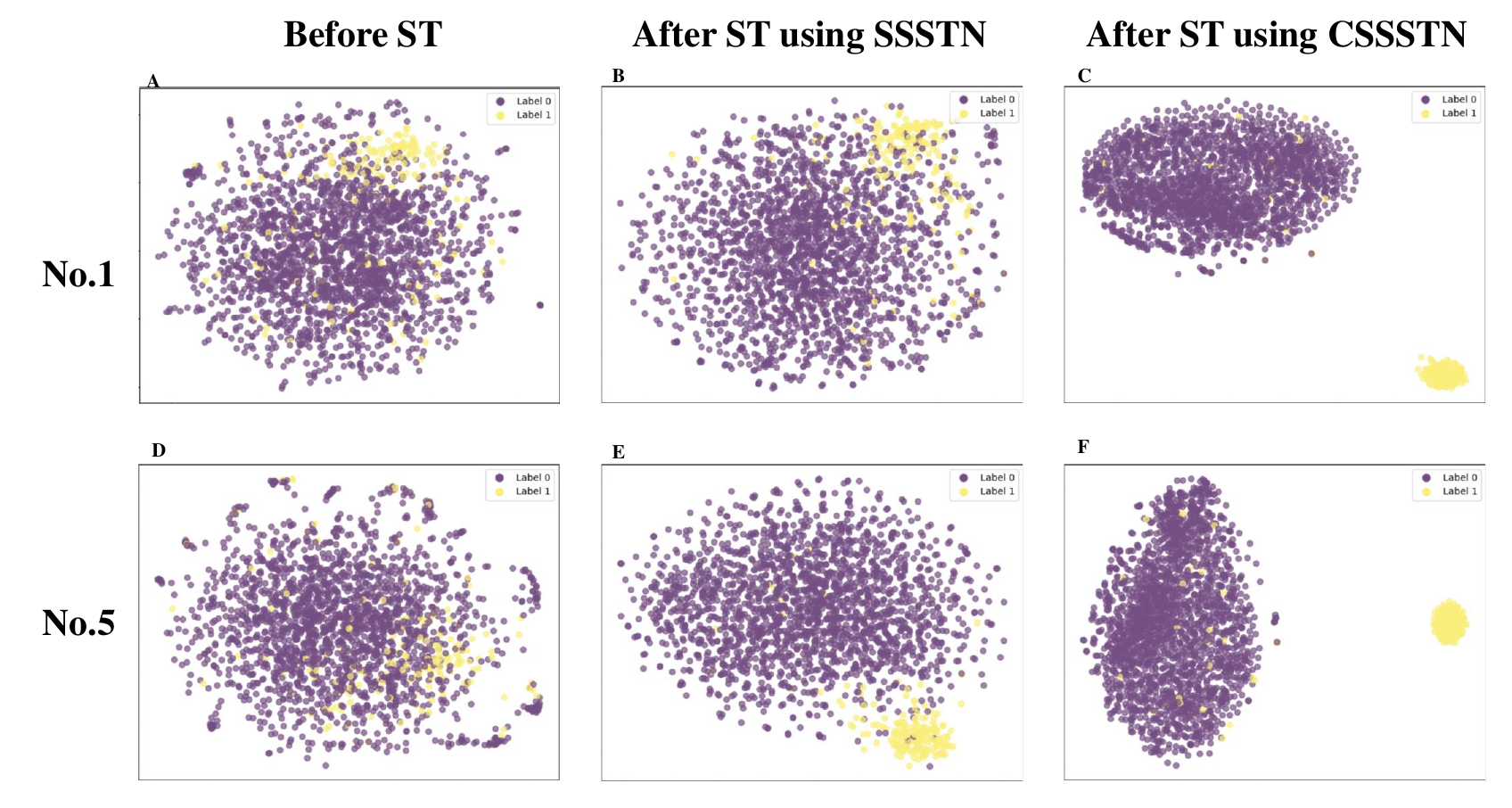}
   \caption{t-SNE visualization of the change in the feature distribution of the target (S01, S05) before and after the style transfer. (A,D) Before style transfer. (B,E) After style tranfer in SSSTN. (C,F) After style tranfer in CSSSTN.}
   \label{fig:tsne}
\end{figure}

To evaluate the effect of style transfer, we applied the t-SNE algorithm \cite{van2008visualizing} to visualize the features \( h \) extracted from the first convolutional layer of \( C \) before and after transformation. This experiment was carried out on the test set of Subjects 1 and 5 in the TsingHua dataset, randomly selected for analysis. Figs.~\ref{fig:tsne}A and 4C show \( h_T \) before style transfer (ST), while Figs.~\ref{fig:tsne}B and ~\ref{fig:tsne}E (using SSSTN) and Figs.~\ref{fig:tsne}C and ~\ref{fig:tsne}F (using CSSSTN) illustrate \( h_S' \) after ST in the two-dimensional embedding space. It can be seen that the features generated by the CSSSTN method show a clear separation between the different classes, providing a solid foundation for the subsequent classification. In contrast, the SSSTN method exhibits a less distinct class boundary in the feature space for S01, which might explain its negative transfer performance. However, for S05, the class boundaries are more clearly defined, indicating better performance for SSSTN in this case.

\subsection{Impact of Target Sample Size}
The size of the target data is critical for calibrating cross-subject EEG models, as fewer samples reduce the calibration time and effort. In this section, we analyze the impact of the target sample size on transfer performance. The results are presented in Table~\ref{table3} for the Tsinghua data set and Table~\ref{table4} for the HDU data set. Different proportions of target data were used for classifier training and style transfer, selected in chronological order (e.g., 25\% refers to the earliest 25\% of collected data).

For both datasets, the baseline classifier models improved consistently as the sample size increased. Surprisingly, CSSSTN did not exhibit the same trend; in some cases, its performance even decreased with more data. This observation indicates that CSSSTN can achieve comparable results using only a small subset of target data, eliminating the need for the full dataset. Thus, our method significantly reduces the calibration time and manual effort while maintaining high performance.

\begin{table}[!h]\tiny
    \renewcommand{\arraystretch}{1.2} 
    \caption{Classification performance on Tsinghua datasets with varying proportions of target data. The highest and second-highest values in each row are highlighted in bold.}
    \label{table3}
    \centering
    \resizebox{\columnwidth}{!}{
        \begin{tabular}{c c c c c c c c}
            \hline
            \multicolumn{1}{c}{\multirow{2}{*}{Subject}} & \multicolumn{2}{c}{25\% data} & 
            \multicolumn{2}{c}{50\% data} & \multicolumn{2}{c}{75\% data} \\ 
            \cline{2-7} 
            \multicolumn{1}{c}{}   & CNN     & CSSSTN   & CNN     & CSSSTN  & CNN  & CSSSTN \\ 
            \hline
             No.1 & 84.2±4.4 & \textbf{94.7±4.4} & 84.9±3.0 & 92.4±2.9 & 85.5±1.7 & \textbf{96.3±4.1} \\ 
             No.2 & 73.3±5.4 & 81.4±5.3 & 77.7±2.2 & \textbf{88.9±4.8} & 79.2±2.1 & \textbf{89.9±4.2} \\ 
             No.3 & 90.8±5.2 & 90.8±5.2 & 91.0±2.5 & 91.0±2.5 & \textbf{91.3±2.4} & \textbf{96.2±2.1} \\ 
             No.4 & 72.7±5.9 & 76.6±4.0 & 73.9±6.2 & \textbf{83.4±2.4} & 78.6±3.1 & \textbf{85.8±5.4} \\ 
             No.5 & 78.1±1.7 & \textbf{95.3±9.2} & 79.5±4.5 & 84.6±2.6 & 83.4±3.7 & \textbf{89.8±3.3} \\ 
             No.6 & 82.7±6.3 & \textbf{98.9±2.0} & 83.7±5.1 & \textbf{97.7±4.6} & 87.6±0.5 & 96.7±3.6 \\ 
             No.7 & 79.4±5.1 & \textbf{90.1±7.6} & 86.3±1.7 & 89.2±4.1 & 87.8±2.3 & \textbf{95.3±2.3} \\ 
             No.8 & 83.2±6.2 & \textbf{94.0±3.7} & 83.1±3.7 & \textbf{95.2±2.5} & 83.4±1.7 & 87.5±2.2 \\ 
             No.9 & 69.0±9.4 & \textbf{86.7±4.7} & 68.4±3.6 & \textbf{78.1±3.4} & 72.1±4.0 & 71.5±3.5 \\ 
             No.10 & 88.6±2.9 & \textbf{95.6±2.2} & 90.4±2.3 & \textbf{99.5±0.6} & 89.2±0.8 & 92.5±1.9 \\ 
             \cline{1-7} 
             Average & 80.2±4.8 & \textbf{90.4±4.8} & 81.9±3.5 & 90.0±3.0 & 83.8±2.2 & \textbf{89.7±3.3} \\ 
            \hline
        \end{tabular}
    }\\
\end{table}

\begin{table}[!h]\tiny
    \renewcommand{\arraystretch}{1.2} 
    \caption{Classification performance on HDU datasets with varying proportions of target data. The highest and second-highest values in each row are highlighted in bold.}
    \label{table4}
    \centering
    \resizebox{\columnwidth}{!}{
        \begin{tabular}{c c c c c c c c}
            \hline
            \multicolumn{1}{c}{\multirow{2}{*}{Subject}} & \multicolumn{2}{c}{25\% data} & 
            \multicolumn{2}{c}{50\% data} & \multicolumn{2}{c}{75\% data} \\ 
            \cline{2-7} 
            \multicolumn{1}{c}{}   & CNN     & CSSSTN   & CNN     & CSSSTN  & CNN  & CSSSTN \\ 
            \hline
             No.1 & 69.6±6.4 & \textbf{92.8±6.9} & 72.2±2.7 & 78.7±3.3 & 75.3±4.1 & \textbf{91.0±5.0} \\ 
             No.2 & 76.3±3.4 & \textbf{95.5±5.5} & 73.0±1.4 & 91.8±5.2 & 75.2±1.2 & \textbf{93.6±8.8} \\ 
             No.3 & 73.3±6.8 & 73.3±6.8 & 74.7±4.2 & 74.7±4.2 & 80.3±2.4 & \textbf{80.3±2.4} \\ 
             No.4 & 67.3±7.0 & \textbf{91.0±4.4} & 73.5±4.2 & \textbf{92.6±7.9} & 75.8±3.1 & 85.5±3.2 \\ 
             No.5 & 81.7±3.3 & 88.2±4.6 & 73.5±2.9 & \textbf{97.4±4.1} & 81.0±2.0 & \textbf{94.9±5.0} \\ 
             No.6 & 75.6±5.8 & 89.5±4.4 & 79.0±4.1 & \textbf{98.0±3.5} & 84.1±6.5 & \textbf{96.7±5.0} \\ 
             No.7 & 68.3±5.2 & \textbf{94.7±9.7} & 68.8±2.6 & 86.2±6.9 & 68.3±4.1 & \textbf{87.7±9.2} \\ 
             No.8 & 73.6±9.3 & \textbf{96.4±6.9} & 79.0±2.7 & 95.6±8.9 & 81.6±7.9 & \textbf{95.8±7.4} \\ 
             No.9 & 62.2±4.9 & \textbf{76.2±3.4} & 60.5±1.5 & \textbf{89.0±8.6} & 69.4±1.8 & 70.7±3.6 \\ 
             No.10 & 65.7±7.7 & 77.1±5.5 & 65.0±1.9 & \textbf{90.1±6.7} & 66.7±1.9 & \textbf{88.1±9.7} \\ 
             \cline{1-7} 
             Average & 71.4±6.0 & 87.5±5.8 & 71.9±2.4 & \textbf{89.4±5.9} & 75.8±3.5 & \textbf{88.5±5.9} \\ 
            \hline
        \end{tabular}
    }\\
\end{table}

\subsection{Ablation Study}

To validate the effectiveness of our proposed CSSSTN method and its loss functions, we conducted an ablation study on the Tsinghua dataset. Table~\ref{table7} outlines the different variants of the CSSSTN model used in this study:
\begin{itemize}
    \item \textbf{CSSSTN w/o \(\mathcal{L}_{\text{cont}}\)}: Excludes the content loss.
    \item \textbf{CSSSTN w/o \(\mathcal{L}_{\text{style}}\)}: Omits the style loss.
    \item \textbf{CSSSTN w/o \(\mathcal{L}_{\text{sem}}\)}: Removes the semantic loss.
    \item \textbf{CSSSTN w/o class}: Excludes class-specific information during style transfer.
    \item \textbf{CSSSTN-A}: Uses all three losses (content, style, and semantic), but calculates the content and style losses using the second-layer features of the classifiers.
    \item \textbf{CSSSTN-B}: Uses all three losses, with content and style losses calculated across all layer features of the classifiers.
    \item \textbf{CSSSTN}: Represents the complete model as proposed in this paper, using the first-layer features for content and style losses.
\end{itemize}

The results demonstrate that all variants of CSSSTN achieved a higher mean accuracy than the CNN model. Each component of CSSSTN positively contributed to the overall performance, with class-sensitive transfer showing particularly significant improvements.

Additionally, among the different approaches to use classifier features during transfer, the best performance was observed when the loss of content and style was calculated using the first layer features. This may be because the lower layer features capture the fundamental patterns necessary for transfer, while the higher layer features may introduce noise or task-specific details that are less effective for transfer learning.

\begin{table*}[!h] \tiny 
    \renewcommand{\arraystretch}{1.2} 
    \caption{Ablation study results demonstrating the impact of removing specific components from the proposed CSSSTN method on the Tsinghua dataset.}
    \label{table7}
    \centering
    \resizebox{\textwidth}{!}{ 
        \begin{tabular}{c c c c c c c c c}
            \hline
            \multicolumn{1}{c}{Subject} & CNN & CSSSTN w/o $\mathcal{L}_{\text {cont}} $   & CSSSTN w/o $\mathcal{L}_{\text {style}}$ & CSSSTN w/o $\mathcal{L}_{\text {sem}} $ & CSSSTN w/o class & CSSSTN-A & CSSSTN-B & CSSSTN\\
            \hline
             No.1 & 85.7±3.2 & 90.3±2.1 & \textbf{93.1±2.4} & 89.5±1.8 & 65.6±4.8 & 92.6±1.4 & 90.3±1.9 & 91.7±1.6 \\ 
             No.2 & 76.9±2.0 & 84.0±3.7 & 84.5±2.3 & 77.8±3.3 & 55.7±2.5 & 86.0±2.4 & 85.6±2.4 & \textbf{86.2±1.1} \\ 
             No.3 & 91.2±1.3 & 93.4±1.0 & 92.8±2.2 & 93.2±1.5 & 64.5±2.7 & 93.1±1.3 & 93.8±1.5 & \textbf{94.8±1.8} \\ 
             No.4 & 77.9±2.1 & 85.9±2.2 & 82.6±3.2 & 81.5±2.4 & 58.4±1.1 & 86.4±2.8 & 85.8±2.7 & \textbf{87.1±3.2} \\ 
             No.5 & 86.2±3.2 & 88.3±3.1 & 86.7±4.6 & 86.8±3.6 & 66.0±3.5 & 86.8±3.4 & \textbf{93.1±3.5} & 92.7±3.4 \\ 
             No.6 & 88.5±1.5 & 91.4±1.5 & 92.6±1.6 & 89.5±1.7 & 65.5±2.4 & 89.9±1.5 & 90.4±1.2 & \textbf{91.5±1.9} \\ 
             No.7 & 89.6±1.8 & 90.1±2.3 & 90.0±3.2 & 88.6±2.0 & 66.6±1.6 & \textbf{93.5±3.6} & 87.8±3.5 & 91.3±3.2 \\ 
             No.8 & 83.7±3.6 & 85.0±4.2 & 84.5±5.2 & 85.4±4.1 & 76.6±7.3 & 85.5±4.2 & 85.3±4.3 & \textbf{93.5±4.3} \\ 
             No.9 & 73.1±3.3 & 85.6±2.9 & 78.5±5.7 & 77.8±4.6 & 63.6±3.0 & 86.6±4.4 & 86.3±4.8 & \textbf{91.4±4.7} \\ 
             No.10 & 91.5±1.3 & 91.5±1.3 & 91.5±1.3 & 91.5±1.3 & 91.5±1.3 & 91.5±1.3 & 91.5±1.3 & 91.5±1.3 \\ 
            \hline
             Average & 84.3±5.3 & 88.6±2.4 & 87.7±3.2 & 86.2±2.6 & 67.4±3.0 & 89.2±2.6 & 89.0±2.7 & \textbf{91.2±2.7} \\
            \hline
        \end{tabular}
    }
\end{table*}

\subsection{Impact of Target Sample Size on Transfer Performance}
An interesting finding from our results is that increasing the number of target samples used for style transfer does not necessarily improve performance. To further investigate whether this phenomenon is related to a decline in the quality of the EEG data over time, we compared the performance of the models trained and transferred using three different subsets of target data: the earliest 25\%, the latest 25\% and a randomly selected 25\% of the data. The results presented for both data sets show no significant differences among these scenarios.This observation suggests that the quality of EEG data does not degrade significantly as the collection progresses and that the performance of our proposed method is not highly dependent on the specific subset of data used for transfer. 

This finding indicates that style transfer may not require large amounts of target data to be effective, as even smaller subsets of the data can yield comparable results. More analysis is needed to explore potential factors that contribute to this phenomenon, such as noise characteristics, subject fatigue, or adaptive neural responses during prolonged EEG data collection. Understanding these factors could help refine the transfer process and enhance the robustness of the proposed method under various experimental conditions.

\subsection{How to Select the Golden Subject}
To explore the impact of golden subject selection on transfer performance, we classified subjects in each data set into two groups: BCI-illiterate subjects and golden subjects, based on their overall classification performance. Tables~\ref{table5} and ~\ref{table6} present the results of CSSSTN's performance when transferring from BCI-illiterate subjects to different golden subjects on the Tsinghua and HDU datasets, respectively. The "Promote" column indicates the improvement in balanced accuracy achieved by CSSSTN compared to the baseline CNN classifier.

\begin{figure}[t]
  \centering
   \includegraphics[width=0.95\linewidth]{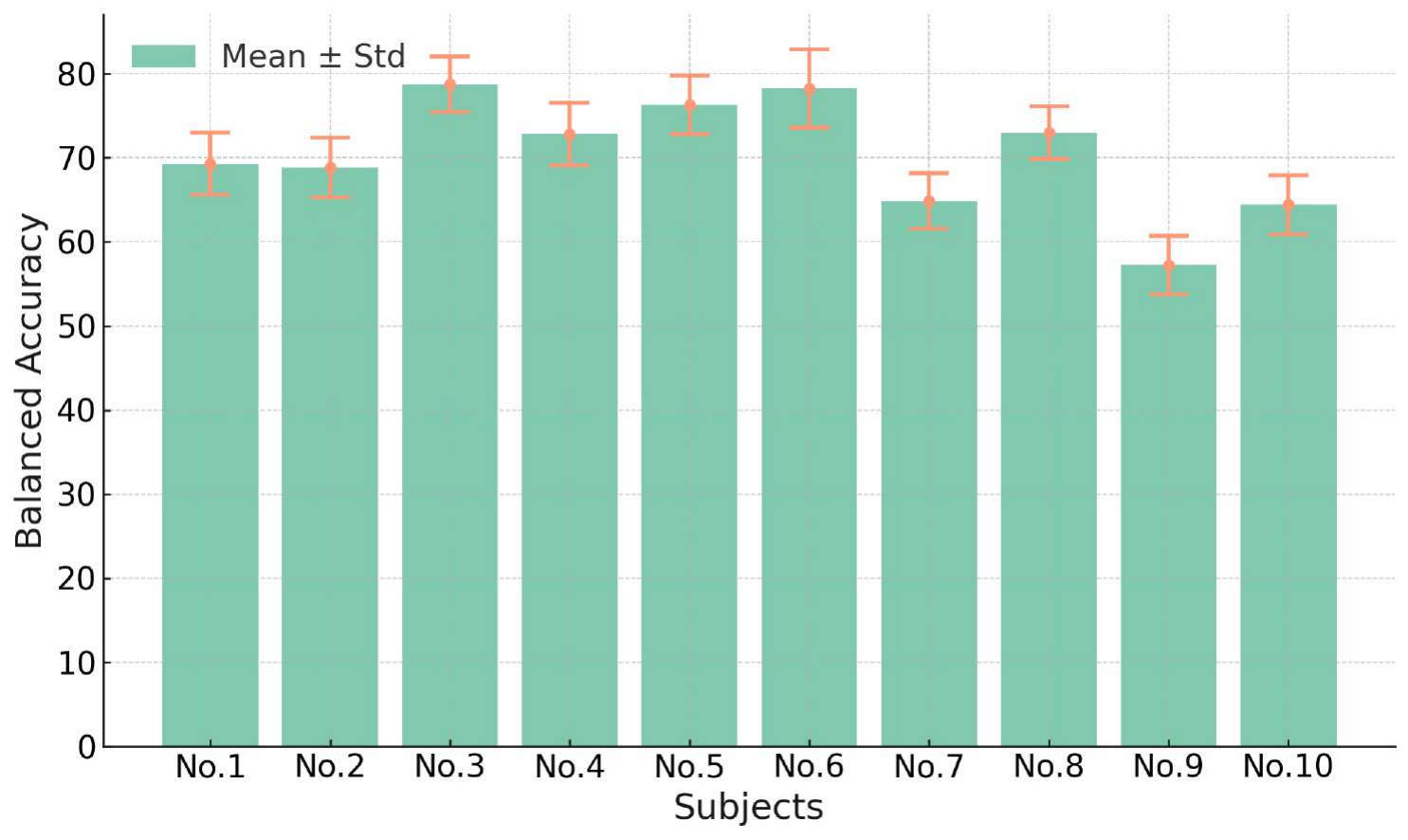}
   \caption{Average balanced accuracy and standard deviation for all 10 subjects across the baseline classifiers (EEGNet, XDAWN-riemann, DeepConvNet, CNN, SE-CNN) in HDU RSVP dataset.}
   \label{fig:different subjects}
\end{figure}

The results show that in the HDU dataset, regardless of subject pairing, CSSSTN consistently outperformed the CNN baseline, with accuracy improvements ranging from 5.8\% to 33.9\%. However, in the Tsinghua dataset, while CSSSTN generally achieved better performance, improper source-target pairings sometimes led to negative transfer. Fortunately, S10 emerged as an undisputed golden subject for the Tsinghua dataset, as it achieved the best classification results in all baseline classifiers (EEGNet, XDAWN-riemann, DeepConvNet, CNN, SE-CNN) (see Table~\ref{table1}). In contrast, identifying a single golden subject in the HDU dataset proved to be more challenging, as different classifiers identified different subjects as the best performing ones. To address this, we averaged the performance of all 10 subjects across the baseline classifiers (EEGNet, XDAWN-riemann, DeepConvNet, CNN, SE-CNN) (as shown in Fig.~\ref{fig:different subjects}) and selected S03 as the golden subject based on its highest average accuracy. Post hoc analysis, as shown in Table~\ref{table6}, confirms that S03 is either the best or second-best option for BCI-illiterate subjects.

Our findings suggest that selecting a golden subject should consider both average performance and standard deviation across multiple baseline classifiers rather than relying on a single classifier. Although this approach may not always yield the best result for every individual subject, it provides a robust and consistent choice overall. This observation is consistent with the findings of previous studies (Biao Sun et al. \cite{sun2022golden}). However, selecting a golden subject based on the results of multiple classifiers can be time consuming. We also explored various distance metrics to measure similarity between subjects, but did not find any significant correlation between similarity measures and the best-performing subjects. Future research could aim to develop a more elegant and efficient method for identifying personalized golden subjects, which could further enhance the practical application of this approach.

\begin{table}[!h] \tiny 
    \renewcommand{\arraystretch}{1.2} 
    \caption{Mean balanced accuracy and standard deviation (std) of BCI-illiterate subjects transferring to different golden subjects on the Tsinghua dataset.}
    \label{table5}
    \centering
    \resizebox{\columnwidth}{!}{ 
        \begin{tabular}{c c c c}
            \hline
            \multicolumn{1}{c}{BCI-illiterates} & Golden subjects & Balanced Accuracy    & Promote \\ 
            \hline
             No.2 & No.1 & 80.5±2.2 & +3.6  \\ 
             No.2 & No.3 & 81.0±2.1 & +4.1  \\ 
             No.2 & No.5 & 72.7±4.1 & -4.2  \\ 
             No.2 & No.6 & 79.7±3.5 & +2.8 \\ 
             No.2 & No.7 & 77.9±1.7 & -1.0  \\ 
             No.2 & No.8 & 73.1±1.2 & -3.8  \\ 
             No.2 & No.10 & \textbf{86.2±1.1} & +9.3 \\ 
             No.4 & No.1 & 84.3±3.9 & +6.4  \\ 
             No.4 & No.3 & 80.6±1.9 & +2.7  \\ 
             No.4 & No.5 & 76.7±3.1 & -1.2  \\ 
             No.4 & No.6 & 83.0±2.3 & +5.4  \\ 
             No.4 & No.7 & 80.2±4.5 & +2.3  \\ 
             No.4 & No.8 & 79.0±2.3 & +1.1  \\ 
             No.4 & No.10 & \textbf{87.1±3.2} & +9.2  \\ 
             No.9 & No.1 & 79.4±5.0 & -6.3  \\ 
             No.9 & No.3 & 83.4±4.6 & +10.3  \\ 
             No.9 & No.5 & 77.3±3.2 & +4.2  \\ 
             No.9 & No.6 & 85.9±6.7 & +12.8  \\ 
             No.9 & No.7 & 83.6±6.1 & +10.5  \\ 
             No.9 & No.8 & 75.8±3.9 & +2.7 \\ 
             No.9 & No.10 & \textbf{91.4±4.7} & +18.3  \\ 
            \hline
        \end{tabular}
    }
\end{table}

\begin{table}[!h] \tiny 
    \renewcommand{\arraystretch}{1.2} 
    \caption{Mean balanced accuracy and standard deviation (std) of BCI-illiterate subjects transferring to different golden subjects on the HDU dataset.}
    \label{table6}
    \centering
    \resizebox{\columnwidth}{!}{ 
        \begin{tabular}{c c c c}
            \hline
            \multicolumn{1}{c}{BCI-illiterates} & Golden subjects & Balanced Accuracy    & Promote \\ 
            \hline
             No.1 & No.3 & 92.1±2.7 & +16.4 \\ 
             No.1 & No.5 & 91.9±5.7 & +16.2 \\
             No.1 & No.6 & 91.6±3.0 & +15.9  \\
             No.1 & No.8 & \textbf{94.2±3.3} & +18.5 \\
             No.2 & No.3 & 95.0±3.5 & +18.1 \\
             No.2 & No.5 & 94.0±1.7 & +17.1 \\
             No.2 & No.6 & 91.8±1.8 & +14.9 \\
             No.2 & No.8 & \textbf{95.6±3.6} & +18.7 \\
             No.4 & No.3 & \textbf{86.1±6.7} & +12.8 \\
             No.4 & No.5 & 85.9±7.0 & +12.6 \\
             No.4 & No.6 & 79.1±3.3 & +5.8 \\ 
             No.4 & No.8 & 81.4±4.0 & +8.1 \\ 
             No.7 & No.3 & \textbf{90.1±8.2} & +17.3  \\ 
             No.7 & No.5 & 90.0±7.7 & +17.2 \\
             No.7 & No.6 & 83.6±6.1 & +10.8 \\
             No.7 & No.8 & 85.8±7.4 & +13.0 \\
             No.9 & No.3 & \textbf{93.6±3.5} & +33.9 \\ 
             No.9 & No.5 & 86.6±1.3 & +26.9 \\ 
             No.9 & No.6 & 82.6±8.7 & +22.9 \\ 
             No.9 & No.8 & 80.8±1.0 & +21.1 \\ 
             No.10 & No.3 & 89.2±8.7 & +17.6 \\ 
             No.10 & No.5 & \textbf{89.7±5.9} & +18.1 \\ 
             No.10 & No.6 & 81.8±6.2 & +10.2 \\ 
             No.10 & No.8 & 84.8±6.1 & +13.2 \\ 
            \hline
        \end{tabular}
    }
\end{table}

\subsection{Challenges and Future Work}
Our proposed ass-sensitive subject semantic style transfer network is a relatively flexible framework. We have not extensively explored the architectures for classifiers and generators, but other advanced approaches can be easily integrated into our network. Additionally, while our work can partially address the cross-subject problem, RSVP-based BCIs still face the challenge of cross-time variations across sessions. Future research could explore the application of style transfer to address cross-session variability. In addition, RSVP-based BCIs also encounter cross-scene issues, where EEG signals may vary depending on the type of target. For example, disguised targets can lead to reduced P300 amplitudes and signal delays compared to clear targets. Transferring the style of EEG signals induced by clear targets to those induced by disguised targets to improve decoding accuracy could be a promising direction for future work.

\section{Conclusion}
In this study, we proposed CSSSTN, a class-sensitive subject semantic style transfer network, to address the challenge of BCI illiteracy in EEG-based RSVP target detection tasks. Building upon the SSSTN framework, CSSSTN incorporates a class-sensitive approach to align feature distributions from source subjects (BCI experts) to target subjects (BCI illiterates) for each class separately. It leverages style loss to perform subject-to-subject semantic style transfer, content loss to capture invisible feature-level semantic styles, and semantic loss to preserve class-relevant semantic information of target subjects. This design addresses both the cross-subject variability and class-specific alignment, which are critical challenges in RSVP-based BCIs.

We evaluated CSSSTN on both a publicly available data set and a self-collected data set. The experimental results demonstrated that CSSSTN outperforms state-of-the-art approaches in mean balanced accuracy, achieving 6.4\% and 3.5\% improvements in the Tsinghua and HDU datasets, respectively, particularly benefiting BCI illiterate users. In addition, t-SNE visualization confirmed the effectiveness of CSSSTN in achieving meaningful feature-level semantic style transfer, showcasing its ability to effectively align class-specific distributions. Furthermore, experiments with varying proportions of target data revealed that CSSSTN achieves comparable or superior performance using only 25\% of the target data, significantly reducing the time and effort required for data collection. This reduction makes the approach more practical for real-world applications by accelerating the training process while maintaining accuracy. We also performed an ablation study to assess the contributions of each component of the model, highlighting the critical role of class-sensitive transfer and integrated losses in improving performance.

In general, this study provides a promising solution for reducing BCI illiteracy and advances the field of subject-to-subject style transfer. By reducing the dependency on large amounts of target-subject data and improving cross-subject generalization, CSSSTN facilitates the transition of RSVP-based BCIs from controlled laboratory environments to real-world applications, enabling broader adoption of BCIs in practical settings. 

\section*{Code Availability}
Code for proposed
CSSSTN model can be found at \underline{https://github.com/ziyuey/CSSSTN}

\section*{Acknowledgement}
This work was supported by National Natural Science Foundation of China (U20B2074, 62471169), Key Research and Development Project of Zhejiang Province (2023C03026, 2021C03001, 2021C03003), Key Laboratory of Brain Machine Collaborative Intelligence of Zhejiang Province (2020E10010), and supported by Zhejiang Provincial Natural Science Foundation of China (No.LQN25F020013).

\bibliographystyle{cas-model2-names}

\bibliography{cas-dc-template}



\end{document}